# Optimal Utility-Energy tradeoff in Delay Constrained Random Access Networks

Amirmahdi Khodaian, Babak H. Khalaj and Hamed Shah-mansouri
Electrical Engineering Department, Sharif University of Technology, Tehran, Iran
Email: khodaian@ee.shrif.edu, khalaj@sharif.edu, hshahmansour@ee.sharif.edu

*Abstract*— **Rate, energy and delay are three main parameters of interest in ad-hoc networks. In this paper, we discuss the problem of maximizing network utility and minimizing energy consumption while satisfying a given transmission delay constraint for each packet. We formulate this problem in the standard convex optimization form and subsequently discuss the tradeoff between utility, energy and delay in such framework. Also, in order to adapt for the distributed nature of the network, a distributed algorithm where nodes decide on choosing transmission rates and probabilities based on their local information is introduced.**

## I. INTRODUCTION

In many wireless ad-hoc networks due to the lack of a central station, nodes compete for the channel and decide channel access in a random manner. In the random access networks that channel access is not predetermined and depends on the network traffic, it is possible that two nodes simultaneously decide to send data to the same node, resulting in collision. Collisions waste energy, increase transmission delay and reduce throughput. The aim of the random access protocols is controlling collisions in the network in order to achieve the desired network performance.

In an earlier paper [1], we solved the problem of energy-utility optimization in random access networks with no delay constraints. In this paper, the notion of delay is added to the analysis of random access networks using the queuing theory. The delay constraint has a non-convex form and in order to convert it into a convex constraint a complete problem reformulation is proposed. An optimal random access protocol which satisfies delay constraints for the links is subsequently presented.

The importance of energy efficiency in ad-hoc networks stems from the multi-hop nature of the network. If nodes of an ad-hoc network run out of energy some routes may become disconnected [2]. Therefore, the available energy of nodes should be consumed cautiously to transmit as much information as possible. Another criterion for the network performance is the utility, which is a monotonically increasing function of the allocated rate to each link. Network Utility Maximization (NUM) has recently received much attention in the literature [5], [6], [7]. It has been first proposed by Kelly [5] in order to optimize end-to-end rates of the wired networks. It is also used in optimizing transport layer of wireless networks [6], [7]. Also, Nandagopal et. al. [8] used similar approach in proportionally fair channel allocation and [9] developed the idea of optimizing persistence probabilities in the random access wireless network. Transmission delay is another important parameter for the network performance and a delay limit should be practically considered for the packets in the network. Such delay constraint depends on the type of traffic and the required quality of service (QoS) level. Real-time applications such as voice or video conferencing require packets to be transmitted with a specific delay limit. However, data transmission is less sensitive to delay and a more relaxed delay constraint can be adopted for such transmissions. It should be noted that one of the issues in random access networks is the size of queues, as for example, in Aloha it is possible that average length of the queues in some nodes go to infinity. Setting a delay limit for the packet transmission is therefore equivalent to setting a limit for the average queue length.

Energy minimization and lifetime maximization for wireless ad-hoc networks have been the focal point of many research activities [10], [11]. However, to the best knowledge of the authors, the work presented in this paper is the first case which considers energy minimization along with delay constraints in random access networks. For example, although [6] and [9] have formulated and solved NUM for the random access but they have neither considered energy consumption nor delay constraints. Optimal utility-lifetime tradeoff has been achieved in [12] for non-random access networks, however no delay constraints were considered in that approach. Delay minimization for slotted aloha was considered in [13] where transmission probabilities were optimized to achieve minimum delay and maximum throughput. However energy minimization and fairness among nodes were not addressed.

The rest of the paper is organized as follows. First, the network model is presented in the next section. Then, in section III we formulate the problem by defining the goal functions and the link delay constraint. Section IV investigates the trade-off between energy, utility and delay; it also contains numerical results of the distributed algorithm. Finally, we conclude the paper and review its contributions in section V.

## II. NETWORK MODEL

Suppose an ad-hoc network which contains $N$ nodes that are going to transmit their packets through their neighbors using the set of links $L$. Each node selects one of its links and transmits with probability $p_{ij}$ where $i$ is the transmitter index and $j$ is the receiver index. $P_i$ is the transmission probability of

This work is supported in part by Advanced Communication Research Institute (ACRI) and Iran Telecommunication Research Center (ITRC). The work of B. H. Khalaj is under support of Alexander von Humboldt foundation.

node $i$, which is equal to the sum of transmission probabilities of its output links. We assume that nodes transmit during time slots whose duration equals the packet transmission time. Collision happens if two neighbors transmit packets in the same slot. Nodes are supposed to have infinite buffers, thus there is no packet drop. We also assume that the distribution of packet arrival at each node is Poisson and independent of the other nodes.

The set of neighbors of the node $i$ is denoted by $N_i$, the set of nodes which $i$ transmits to them with $O_i$ and the set of nodes that transmit to $i$ with $I_i$. We define connectivity factor by the ratio of communication range to the network dimension. Thus, as the nodes' powers increase, the connectivity factor and number of neighbors, $|N_i|$, increase as well. In this paper, we assume that all nodes have equal power, resulting in symmetric neighborhoods. The case that neighbors use unequal powers was considered in [1]. Although such assumption can be easily incorporated in the current work, it has not been considered in this paper in order to simplify the formulations.

## III. DELAY ANALYSIS

In order to calculate average delay in random access networks, we assume packet arrival at each node to be modeled by a Poisson process. It is also assumed that in case of collisions, the packet is retransmitted until it is successfully received at the other end. Thus, when a packet collides it does not return to the queue, but waits until it is served. The service time of each link depends on the transmission probability of that link and the collision probability. We can model each link as an *M/G/1* queue and use the following Pollaczek-Khinchin formula to estimate the queue delay [14].

$$T = W + \overline{S} = \overline{S} + \frac{r\overline{S^2}}{2(1-\rho)} \qquad (1)$$

where $W$ is the waiting time of the queue, $\overline{S}$ is the average service time, $r$ is the arrival rate, and $\rho = r\overline{S}$. Thus, the first and second order mean of the service time should be computed. In the slotted access, the service time is a discrete random variable and the probability of transmission after $k$ time slots is equal to $x(1-x)^k$, where $x$ is the probability of successful transmission. Mean and variance of the service time are then given by:

$$\overline{S} = 1/x$$
$$\sigma_S^2 = \frac{1-x}{x^2} \qquad (2)$$

Thus, using (1) and (2) the link delay can be found as follows:

$$T = \frac{1}{x} + \frac{r\frac{2-x}{x^2}}{2(1-\frac{r}{x})} = \frac{1}{x} + \frac{r(2-x)}{2x(x-r)} = \frac{(1-r/2)}{(x-r)} \qquad (3)$$

## IV. OPTIMAL MAC WITH DELAY CONSTRAINT

Our goal is to optimize MAC parameters in order to achieve minimum energy consumption and maximum utility in the network. Solving such a bi-criterion problem is equivalent to finding Pareto optimal points [15]. Pareto optimal points have the characteristics that no other point that is better in both energy and utility exists. Also, an additional delay limit for the links of the network should be considered in this bi-criterion problem. In problems where goal and constraint functions are convex, it is common to use scalarization in order to find the Pareto optimal points. However, in this case delay constraint in its original form is non-convex and the first step in the proposed algorithm converts it to a convex function. Subsequently, scalarization can be used in order to form a convex problem and achieve Pareto optimal points.

### A. Convex Formulation

The utility function, $U$, is defined as the summation of link utilities. In order to achieve proportional fairness between links we use the same approach as [8] and [9], and define utility as a logarithmic function of the link rates:

$$U = \sum_{(i,j) \in L} \log(r_{ij}) \qquad (4)$$

Therefore, the utility which is the first goal function, is a concave function. Another goal function that should be formulated is the energy consumed in the network. The required energy to transmit a packet by node $i$ is equal to $e_i$, so average energy consumption by node $i$ in one timeslot is given by $E_i = e_i \times P_i$. In this paper we assume equal transmission power for the nodes and thus, total energy consumption of the network is given by the following linear function:

$$E = \sum_{i \in N} E_i = e \sum_{i \in N} P_i \qquad (5)$$

The network parameters targeted to the optimization problem are transmission probabilities and rates. The energy and the negative of the utility are goal functions that should be minimized and were shown to be convex functions of transmission rate and probabilities. The next step is to show that the constraints are also convex functions of these parameters. Link delay constraint is found in section III and can be reformulated as follows:

$$D_{ij} = \frac{(1 - r_{ij}/2)}{(x_{ij} - r_{ij})} < D_c \Rightarrow r_{ij} + \frac{(1 - r_{ij}/2)}{D_c} < x_{ij} \qquad (6)$$

where $x_{ij}$ is equal to the throughput of packets on link $(i,j)$. Using the collision model, successful reception probability depends only on transmission probability of $j$'s neighbors. Therefore, a packet is received successfully if and only if neither $j$ nor any of the neighbors of $j$ except $i$ have sent a packet at the same time. Thus, throughput of a link is given by the multiplication of successful reception probability and link capacity:

$$x_{ij} = c_{ij} p_{ij} (1 - P_j) \prod_{l \in N_j^m - \{i\}} (1 - P_l) \qquad (7)$$

Equation (7) shows that $x_{ij}$ has a product form and is non-convex, so, the delay constraint (6) is non-convex. In order to obtain a convex delay constraint as a function of $p_{ij}$, we first use a logarithmic function, which is monotonically increasing and preserves inequality, on both sides of (6):

$$\log(r_{ij} + \frac{(1-r_{ij}/2)}{D_c}) - \log(c_{ij}p_{ij}) + \log(1-P_j)$$
$$+ \sum_{l \in N_j^{in}-\{i\}} \log(1-P_l) < 0 \quad (8)$$

It is easy to show that the above constraint is a concave function of transmission rates $r_{ij}$. Therefore, by using a change of variables of the form $z_{ij}=\log(r_{ij})$, the delay constraint can be converted into a convex function of $z_{ij}$. This also changes utility to a linear function:

$$U = \sum_{(i,j) \in L} z_{ij} \quad (9)$$

We can now use scalarization for the convex goal functions and constraints in order to formulate a convex problem and find Pareto optimal points:

$$\begin{aligned}\min \quad & \lambda_1 E - \lambda_2 U \\ S.t. \quad & \log(\frac{1}{D_c} + e^{z_{ij}}(1-\frac{1}{2D_c})) - \log(x_{ij}) \leq 0 \\ & 0 \leq P_i, p_{ij} \leq 1 \, \forall i \in N, j \in O_s \\ & P_i = \sum_{j \in O_i} p_{ij} \quad \forall i \in N \end{aligned} \quad (10)$$

There are many well-known algorithms for solving such convex problems. We use Sequential Quadratic Programming (SQP)[16] to solve (10). In section VI, optimal tradeoff curves for energy and utility with different delay constraints are obtained using SQP.

### B. Feasibility of the problem

The convex formulation ensures that the problem has a unique solution in its feasible region. One remaining problem is the issue of feasibility of the problem. This depends on the link delay constraint ($D_c$). It is apparent that using delay constraints smaller than the average service time of any link may turn the problem into an infeasible problem. So, we should find the minimum delay constraint (MinD$_c$), that ensures feasibility of the problem, and only adopt higher delay constraints for the network. The delay constraint formula (6) shows that the maximum link delay occurs for the link with minimum throughput. As a result, if the minimum throughput is maximized over all links, it is possible to obtain the point that can tolerate the MinDc. This is equivalent to the following maxmin optimization problem:

$$\begin{aligned}\max_p \min_{(i,j)} x_{ij} & \\ 0 \leq P_i, p_{ij} \leq 1 & \quad \forall i \in N, j \in O_s \end{aligned} \quad (11)$$

This problem can also be formulated as the convex optimization form of (12). The achieved minimum delay constraint depends only on the network structure. In section V (12) is used to calculate MinDc for different network structures and sizes. Henceforth, we assume that structure of the network is approximately known and $Dc$ in (10) is set so that it is feasible.

$$\begin{aligned}\max_p & \; z \\ & z \leq \log(x_{ij}) \\ & 0 \leq P_i, p_{ij} \leq 1 \quad \forall (i,j) \in L \end{aligned} \quad (12)$$

### C. Distributed MAC Optimization

In general, algorithms such as SQP are applied in a centralized manner. In practice we should use distributed algorithms in the network so that nodes can decide and select their optimal variables through minimum interaction with other nodes. Since, the problem is convex and feasible, the duality gap is zero and we can use the dual problem in order to make separate problems over the nodes. This dual decomposition approach will give update formulas for link probabilities and rates.

First, we write the Lagrangian of (10) as follows:

$$\begin{aligned}L(\mu,p,z) = \sum_{(i,j)} [\lambda_1 e_i p_{ij} - \lambda_2 z_{ij} + \\ \mu_{ij}(\log\left[(1-\frac{1}{2D_c})e^{z_{ij}} + \frac{1}{D_c}\right] - \log(x_{ij}))] \end{aligned} \quad (13)$$

where $\mu_{ij}$ is the dual variable for delay constraint of the link $(i,j)$. Using the derivative of the Lagrangian we can find the rate update formula and the corresponding equation for the link probabilities:

$$r_{ij}^{(n+1)} = \left[\frac{\lambda_2}{(\mu_{ij}^{(n)} - \lambda_2)(D_c - 0.5)}\right]^+ \quad (14)$$

$$p_{ij}^{(n+1)}\lambda_1 e = \mu_{ij}^{(n)} - \frac{1}{(1-P_i^{(n+1)})} p_{ij}^{(n+1)} \sum_{\substack{(k,l) \in L \\ l \in N_i \cup i}} \mu_{kl}^{(n)} \quad (15)$$

Using (15) and computing the summation over $j$, will give us the following quadratic equation for the node probabilities:

$$(P_i^{(n+1)})^2(\lambda_1 e_i) - P_i^{(n+1)}(\sum_{\substack{(k,l) \in L \\ l \in N_i \cup i}} \mu_{kl}^{(n)} + \sum_{l \in O_i} \mu_{il}^{(n)} + \\ + \sum_{l \in O_i} \mu_{il}^{(n)} = 0 \quad (16)$$

New link probabilities can be found using updated node probability and dual variables:

$$p_{ij}^{(n+1)} = \text{proj}\left[\mu_{ij}^{(n)}\left(\lambda_1 e_i + \frac{1}{1-P_i^{(n+1)}}\sum_{\substack{(k,l) \in L \\ l \in N_i \cup i}} \mu_{kl}^{(n)}\right)^{-1}\right] \quad (17)$$

where proj[·] function projects transmission probabilities in the feasible region.

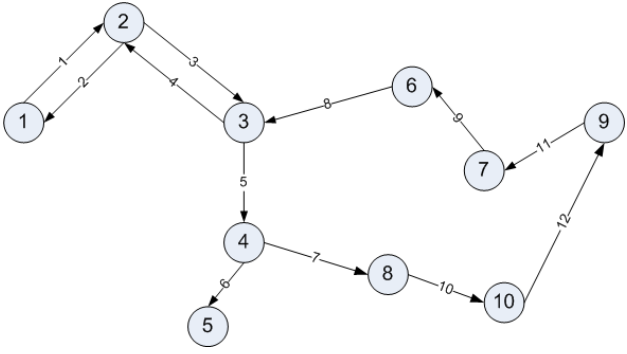

Figure 1.  Sample network

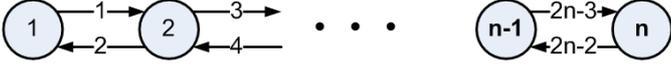

Figure 2.  Linear Network with *n* nodes

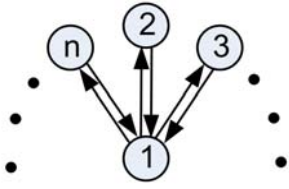

Figure 3.  Star Network

Also, the following formula can be used to update dual variables:

$$\mu_{ij}^{(n+1)} = [\mu_{ij}^{(n)} + \alpha_n \{\log\left[(1-\frac{1}{2D_c})r_{ij}^{(n)} + \frac{1}{D_c}\right] - \log(x_{ij}^{(n)})\}]^+ \quad (18)$$

Convergence of this dual decomposition algorithm is guaranteed for small values of $\alpha_n$ or when $\alpha_n$ goes to zero for large $n$ [15]. In the numerical analysis, we have used a constant small step size since such choice does not require synchronous update of the step size in the whole network and also reduces the complexity.

## V. NUMERICAL ANALYSIS

Three types of networks are considered in our numerical analysis. First, the sample network of Fig. 1 which has 10 nodes and 12 links is considered. Linear and star networks are also used in order to investigate the problem for different network sizes (Fig. 2 and Fig. 3). For the linear network we assume that nodes are only neighbors of the nodes that they have a common link with. In the star network we assume that all nodes are neighbor of node 1 and their adjacent nodes so in Fig. 3 node 2 is neighbor of 1, 3 and *n*. Also, in the numerical analysis we assume $c_{ij}=1$.

### A. Centralized Solution

Minimum delay constraint ($MinD_c$) of the links is a parameter that should be properly set to guarantee the feasibility of the problem. It was shown in section IV.B that this constraint can be found by solving (12). For the sample network of Fig. 1, the *MinDc* that can be used is equal to **10.47**.

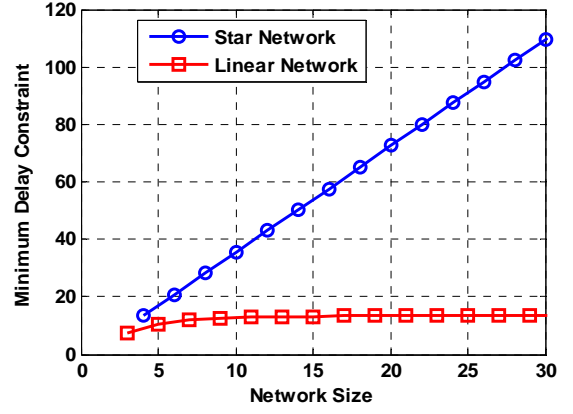

Figure 4.  The minimum delay constraint for star and linear networks

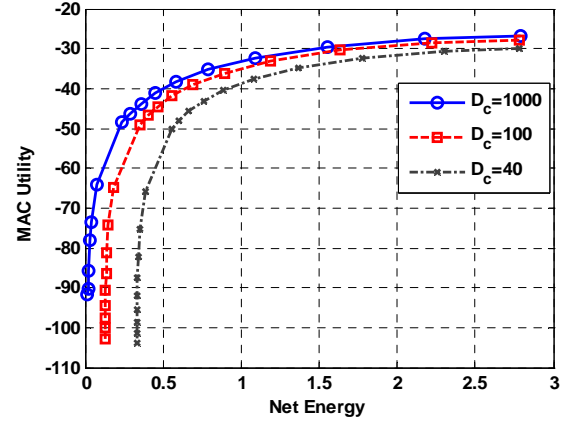

Figure 5.  The Optimal energy utility tradeoff in the sample network for different delay constraints

For linear and star networks the *MinDc* may vary with the network size. As shown in Fig. 4, for linear networks this minimum delay changes very slowly with the network size. However, for star networks it linearly increases as the number of nodes increases.

The cost function of problem (10) is a linear combination of energy and utility. The parameters $\lambda_1$ and $\lambda_2$ can be changed in order to control the tradeoff between energy minimization and utility maximization. Although it is possible to use the *MinDc* given in problem (10), using *MinDc* results in only one feasible point. We use delay constraints of about 4× *MinDc* and more in order to obtain a large enough feasible region. This allows $\lambda_1$ and $\lambda_2$ to better control the tradeoff between energy and utility.

Fig. 5 shows the trade off between energy and utility for three different delay constraints in the case of the sample network. As the delay constraint decreases, the optimal points have less energy and more utility. Our numerical analysis shows that increasing delay from 40 to 100 is more effective than increasing it from 100 to 1000. Also, three regions from left to right can be distinguished on each curve. At large values of $\lambda_1$, the energy is near its minimum value and changes very slowly, but the utility decreases at a high rate. In the next region, the tradeoff between energy and utility is more evident. The last region is where utility slowly reaches its maximum value at the cost of doubling energy consumption from 1.5 to 3.

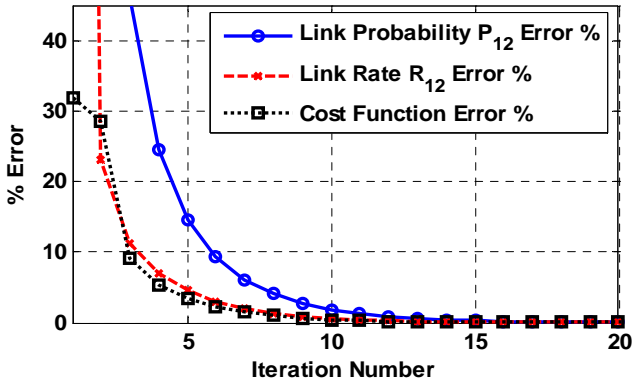

Figure 6. Convergence of the distributed algorithm for the sample network.

## B. Distributed Algorithm

A simple distributed algorithm was described in section IV.C. We have used this algorithm for the case of $(\lambda_1, \lambda_2) = (5, 0.1)$ and a delay constraint equal to 100. The algorithm starts from initial transmission probability of 0.1 for all links. Results of the distributed algorithm are then compared with the optimum point in Fig. 6 where the percentage of error in network cost function, transmission probability of link (1,2), and link data rate is shown. In these curves, the error percentage of the $x$ is defined as $|x(itr) - x^{opt}|/|x^{opt}|$. If we use error of less than 1% as a measure of convergence, it can be verified that the distributed algorithm converges in about 12 iterations for the sample network.

One interesting question to address is how the convergence of this algorithm scales with the network size. In order to investigate such effect, a linear network in which the minimum delay does not scale with the network size has been considered. The convergence rate is then compared for different network sizes in which $D_c=100$ and $(\lambda_1, \lambda_2)=(5,0.1)$. Our numerical analysis shows that for all linear network sizes between 4 and 32, the number of iterations required for convergence is roughly 15. In order to explain this we note that in gradient direction methods computations scale with the dimension of the problem. However, when distributed computation is used, the number of computers also scales with the network size. Consequently, the number of iterations computed by all nodes in finding the optimum point does not scale with the network size.

## VI. CONCLUSION AND FUTURE WORKS

In this paper, the delay constraint was added to the previous work on energy-utility optimization in random access networks. We have modeled links as M/G/1 queues and used this model in order to calculate average delay of the random access protocol. Non-convexity of the delay constraint was the main obstacle for the centralized or distributed optimization algorithms. After proper transformations, the problem is transformed into the convex form. Subsequently, the bi-criterion problem of energy-utility maximization with delay constraint is formulated as a standard convex optimization problem and dual decomposition is used to achieve a distributed solution. This convex problem is not only useful for network design but can also be used to find optimum achievable energy and utility values at different delay constraints. The minimum delay constraint that ensures feasibility of the problem is also considered in the paper. It is shown that a maxmin problem should be solved in order to find MinDc. A convex equivalent is also given for this maxmin problem.

Our numerical analysis shows the trade off between energy, utility and delay in the random access network. We indicated that there are some regions that there is gain for energy or utility only at the cost of loss for the other one. Also increasing the delay constraint near MinDc is more effective than increasing it at large delay values. The convergence rate was another parameter considered in the numerical results. The relationship of the network size and convergence rate was also addressed for linear networks.

The next step in continuation of the current work is to consider delay in the cross-layer problem of MAC-Transport optimization where we should consider end-to-end delays and rates. In this case, the assumption of Poisson arrivals at the nodes should be revised. Our initial work shows that this cross-layer problem is non-convex and non-separable.